# How Does COVID-19 impact Students with Disabilities and Health Concerns?

HAN ZHANG[1], PAULA NURIUS[1], YASAMAN SEFIDGAR[1], MARGARET MORRIS[1], SREENITHI BALASUBRAMANIAN[2], JENNIFER BROWN[1], ANIND K. DEY[1], KEVIN KUEHN[1], EVE RISKIN[1], XUHAI XU[1], JEN MANKOFF[1], [1]University of Washington [2]SASTRA Deemed University, USA

The impact of COVID-19 on students has been enormous, with an increase in worries about fiscal and physical health, a rapid shift to online learning, and increased isolation. In addition to these changes, students with disabilities/health concerns may face accessibility problems with online learning or communication tools, and their stress may be compounded by additional risks such as financial stress or pre-existing conditions. To our knowledge, no one has looked specifically at the impact of COVID-19 on students with disabilities/health concerns. In this paper, we present data from a survey of 147 students with and without disabilities collected in late March to early April of 2020 to assess the impact of COVID-19 on these students' *education* and *mental health*. Our findings show that students with disabilities/health concerns were more concerned about classes going online than their peers without disabilities. In addition, students with disabilities/health concerns also reported that they have experienced more COVID-19 related adversities compared to their peers without disabilities/health concerns. We argue that students with disabilities/health concerns in higher education need confidence in the accessibility of the online learning tools that are becoming increasingly prevalent in higher education not only because of COVID-19 but also more generally. In addition, educational technologies will be more accessible if they consider the learning context, and are designed to provide a supportive, calm, and connecting learning environment.

## 1 INTRODUCTION

The impact of COVID-19 on students in general can potentially be large. Worries about fiscal and physical health, rapid change, and increased isolation are all factors that can affect students and general. Students with disabilities/health concerns may face greater risks than their non-disabled peers in a wide range of aspects. First, they have higher financial risk due to the need for accessible housing and personal assistance. Moreover, they face greater health risks, depending on their specific situation. In addition, the risk to their education is also higher due to potentially inaccessible course content and classroom environments [21, 31]. These risks may increase further in the face of a global health pandemic. An understanding of the impact of COVID-19 on students with disabilities/health concerns is important so that we can identify these increased risks early and respond to them.

In this paper, we define disability broadly to include anyone who self-identifies as disabled and also those with health concerns who may not identify as disabled [15]. We focus on two specific issues of concern to all students: Education and Mental health. Both are currently under stress, and understanding how and whether those stresses differentially impact students with disabilities/health concerns is critically important. By comparing data about students' experiences during COVID-19 we hope to identify areas of concern.

To this end, we present descriptive data showing how students with disabilities/health concerns, in comparison to their non-disabled peers, are faring in the COVID-19 context. We present an analysis of data from a multi-year study at a large public university that has included approximately 200 students per year, including students with and without disabilities/health concerns [36]. In this paper, we focus specifically on the impact from survey results from Spring 2020 (during the COVID-19 pandemic), in which we ask 147 students about their physical and mental health, their academic and social experiences, and specific concerns about COVID-19. Our sample includes 28 students with disabilities/health concerns in 2020. Where relevant, we also make comparisons to data collected from a subset of 91 students, of whom 13 have disabilities/health concerns who filled out a very similar survey in Spring 2019 (before COVID-19).

We contribute insights about university students' early responses just before and during COVID-19, to characterize patterns of concern, challenging experiences, and stress as well as initial indicators of mental health. Our central focus is on students with disabilities/health concerns, and the impact of COVID-19 on their education and health.

*Educational Impacts.* The shift to online learning may impact the types of access technologies and techniques that students use, as well as their dependence on online course materials that may not be accessible [21, 31]. Particularly, as many universities made a late decision to teach online, instructors had less time than usual to prepare materials for an online setting, which may have impacted the accessibility of those materials. At the same time, for some students the ability to participate online may *increase* accessibility. *We contribute an analysis of educational concerns and how they differ from before to during COVID-19.* We compare students with and without disabilities/health concerns, and we analyze differences between different classes of disabilities/health concerns.

*Our findings suggest that students with disabilities/health concerns have increased concerns about classes going online over those without, particularly the impact on their degree goals such as admission to major and graduation timing.*

*Impact on Mental Health.* When social systems are placed under stress due to sudden forces such as a global health pandemic, the impact on marginalized populations can be especially severe. Outbreaks of widespread infectious diseases, such as SARS and COVID-19, are associated with psychological distress and eroded mental health, stimulating calls for mental health care to be part of the national public health emergency system [4]. Our examination of mental health impacts starts with exposure to stressors. We look at several potential sources of stress, including the use of social media, income loss, quarantine and isolation, loneliness, chronic discrimination, and overall exposure to adversity.

At the time our data was collected, just after classes went online and three weeks after it was discovered that community spread of COVID-19 was present in Seattle (Table 1), *we did not find evidence of changes in anxiety, stress, or depression among students with or without disabilities/health concerns. However, students with disabilities/health concerns reported more stress exposure, and more distress, than those without.*

*Implications for Accessible Online Education.* Although our sample is small and the timing early in the COVID-19 crisis, this data provides an important snapshot into the consequential nature of students' concerns about online education, which includes the impact not only on grade, but also on admission to major and graduation.

The COVID-19 epidemic's impact on learning should be a wake-up call to accessibility researchers to study online learning technologies and their impacts, and higher education in general, from a disability perspective. Although social distancing may fade into memory, it is likely that online learning will not. Even before COVID-19, online education was beginning to achieve parity with in-person education [27].

Yet this is a topic that has received little attention from computer scientists or accessibility researchers. A title word search for "education OR course OR teaching" within the ASSETS proceedings found only 29 matches, of which most discuss introducing curricula that would educate technologists about accessibility (*e.g.,* [20]). While some deal with online educational tools in general (*e.g.,* [8] [12]), or childrens' access to the mainstream education system [9], there is much less work published about higher education. Although this is a limited search, it highlights the relative lack of attention to online learning, and higher education in general, in the accessibility community (in contrast, the keyword "blind" found 139 matches).

Our work provides motivation for the importance of improving the accessibility of online classes (and relatedly the ease with which instructors can make such classes accessible). In addition, as Ringland *et al.* argue [33], accessibility solutions must take a holistic, contextualized view of the person, including consideration of their potential stresses and concerns. Thus, we argue that under COVID-19 (and, really, in general) an accessible education is also one that contributes to student well-being and resilience. Our data help to justify this need by demonstrating that, at least in our sample, students with disabilities are operating under increased stress loads both over their lifetime and specifically since the beginning of the COVID-19 pandemic.

## 2 RELATED WORK

Life as an undergraduate student presents an exciting and challenging opportunity to learn, innovate, and grow. However, exposure to stress is also common among college students [18]. In addition, some stressors differentially impact vulnerable groups, and these impacts can translate into changes in mental health and behavior (*e.g.,* [36]), particularly when they interact with risk [35]. Increasing participation by people with disabilities requires understanding and efforts to lessen the particular stresses and risks they face. The COVID-19 pandemic is one obvious stressor that may directly impact students in a multitude of ways. Some may be at greater risk of contracting the illness and interrupted access to health services [3]. Pandemics can also impact availability of personal assistance, reduce educational accessibility [21], and create higher risks of exposure than for the overall population [5]. Below we explore expected consequences for learning and mental health of these stressors for people with disabilities.

### 2.1 The Impact of COVID-19 on Education

One significant impact of COVID-19 has been the shift to online learning. Online learning has become increasingly common over the last decade [31], and even before the radical changes imposed by social distancing in the era of COVID-19, issues existed with the accessibility of online courses. For example, at one university, 75% of faculty reported having never made accommodations in their online material [31], most having never been asked by students to do so. Student reluctance to ask for accommodation is driven in part by concerns about faculty bias [21].

Due in part to the lack of perceived need among professors, the perspective of students with disabilities indicates that online learning is not fully accessible [21]. In addition, the move to increased asynchronous learning can leave students with disabilities behind, as, for example, videos are often not captioned. As of 2015, only Blackboard (of all online learning management systems) had been awarded Gold certification for its accessibility by the National Federation for the Blind [21]. In addition, content generators (*i.e.* faculty) have a big impact on the accessibility of their content [21, 22] and over half (65%) of faculty are unsure or do not know how to make accommodations [31]. This results in many courses being designed and deployed without accessibility support [21]. When faculty additionally have to make very rapid changes to their courses due to pressure to go online quickly, it is even less likely that they will make time to attend to an issue they are unsure of, like accessibility.

At the same time, a shift to online learning due to a pandemic could benefit students with disabilities. For example, going online might reduce the need for disclosure, increase flexibility if instructors are being especially accommodating under the circumstances, or reduce barriers for students with mobility-related impairments.

### 2.2 The Impact of COVID-19 on Mental Health

Even before the pandemic, universities were seeing high and growing levels of mental health struggles among students [17]. Further, natural disasters as well as past epidemics such as SARS, have been associated with psychological distress, depression and substance abuse [13, 16, 23, 28, 41]. Adverse psychological effects of the current pandemic may also be serious, but are not expected to be uniform [30]. For example, many individuals show resilience in disasters; this resilience manifested in a recent study of college students who, contrary to expectations, reported less loneliness in April than in February of this year [10]. Daily surveys by Kanter *et al.* of the general population from March 14th through May 5th 2020 (during the COVID-19 epidemic) have not shown overall increases in anxiety, depression or loneliness (https://uwcovid19.shinyapps.io/dashboard/). On the other hand, students with a history of stress exposure may have raised allostatic load [25], lowering the physiological resilience to the additional demands brought on by the pandemic.

It is unclear whether students with disabilities/health concerns are likely to experience mental health consequences of COVID-19's impact on society. What is expected is that people with disabilities/health concerns are likely to experience more stressors associated with the pandemic. In addition to the risks of COVID-19, and educational accessibility, emergency response plans do not always include planning for people with disabilities [5], social distancing may be harder for people with disabilities who rely on caregivers, and people with disabilities may have to deal with inaccessible communications [5]. These risks may put people with disabilities at greater risk of experiencing distress associated with the pandemic. Their concerns about personal safety may be stronger than those of people without disabilities and they may be more vulnerable to stressors during the pandemic (such as disruption of critical health care). For these reasons they may be at greater risk of psychological distress.

### 2.3 Summary and Research Questions

Given the literature thus far, there is cause for concern. The COVID-19 context appears to be exacerbating risks related to mental health, educational outcomes, and physical health, particularly for students with disabilities. Thus, this paper addresses time-sensitive questions regarding the adversity profiles and well-being of university students during the current pandemic, with particular focus on ways that students with disability needs may differ from nondisabled students. Drawing from research related to cumulative disadvantage and marginality [29], we anticipate that students with disabilities will express justifiably greater vulnerability in the COVID-19 context through elevated levels of concern about their educational environment and likelihood of success, as well as features of their family and personal lives affected by the pandemic. Finally, we anticipate that students with disabilities may be experiencing greater levels of psychological distress in the form of mental health indicators.

## 3 METHODS

Our data are derived from a multi-year study, the University of Washington Experience (UWEXP) [36], which is in its third year of collecting data about college students' behaviors, mental health, education and well-being. This, in turn, was inspired by a study first pioneered by Wang et al. [42]. The goals of the UWEXP study are to understand the stressors impacting undergraduate college students through a combination of self-reported information about demographics, health and well-being, institutional data about educational outcomes, and behavioral data collected using FitBits and mobile phones. In 2020, we recruited from all past participants in the UWEXP study and additionally advertised to the entire University of Washington entering first-year student class. In addition, we conducted targeted recruitment in communities that might be marginalized including students with disabilities, and first-generation college students. Students are paid to participate in the study, which is IRB approved.

In this paper, we primarily make use of the UWEXP survey data collected in Spring of 2020, within the COVID-19 context. Students were instructed to complete the survey after their last Winter Quarter final. At that time, the University of Washington had been teaching online for at least two weeks, and students knew that half or more of Spring quarter would be online. The UWEXP baseline survey data was conducted from March 18th through April 8th (end of Winter quarter through beginning of Spring quarter). The majority (80%) of students completed it between March 18th and March 29th. At the time, social distancing restrictions were increasing city and state-wide, with the initial stay-at-home order for Washington issued March 23rd. The full timeline is shown in Table 1.

Students answered an hour-long questionnaire consisting of demographics and a series of well-established scales to measure depression (CES-D [32]); perceived stress (PSS [7]); anxiety (STAI [39]); loneliness (UCLA [34]); chronic discrimination and harassment (CEDH [26]); Perceived Social Status (SES, measured using the MacArthur Scale of Subjective Social Status [1]) and major life events (MLE [36]). Several of these scales ask students to reflect over a period of time: The depression scale items measure the past week; whereas perceived stress and trauma symptomology assessed the past month. Students reported Major Life Events (adverse, traumatic or stressful events) over their lifetime. They reported their depression-related feelings that occurred during last week. We measured trait anxiety and loneliness does not provide a timescale, it is thus likely that answers to the loneliness scale reflect current (in-the-moment) status.

In addition, since one arm of the UWEXP study was concerned with adversity exposure and related symptoms, the data collection protocol includes a scale designed to measure post-traumatic stress disorder (PTSD) symptomology [43]. We note that in a population without known exposure to severe trauma (such as most of our participants), the items in this scale are better conceptualized as a measure of *distress* rather than indicative of clinical post-traumatic stress.

When no established measures were available, we created questions to address variables of interest such as adverse events due to COVID-19. This included questions about students' levels of concern about current life issues–both educational and family related; questions about COVID-19 specific adversity exposures; and a single item asking if they have a pre-existing condition that makes them vulnerable to COVID-19. In addition, the UWEXP study includes a student-specific measure of COVID-19 related adversities.

| Date | Event |
|---|---|
| 2/28 | Evidence of community spread discovered in Seattle |
| 3/3 | First COVID-19 related death discovered in Seattle |
| 3/6 | Announcement that classes would officially switch to online |
| 3/13 | Last day of instruction for Winter quarter and announcement that Spring quarter will begin online |
| 3/18 | Announcement that Spring quarter will be fully online |
| 3/18 | *Earliest date a student took the baseline survey* |
| 3/20 | Last day of final exams for Winter quarter |
| 3/30 | Instruction for Spring quarter begins |
| 4/8 | *Latest date a student took the baseline survey* |

Table 1. Timeline of announcements and events of relevance at the University of Washington

## 3.1 Data Preparation and Analysis

152 students participated in the 2020 survey, of whom 147 stayed in the study (96.7% retention rate). We removed the data for the five students who dropped out from our data analysis. We assign disability status primarily based on self-identification. Additionally, we added people who gave very high responses (*moderate* or *severe*) on at least eight of the 33 items of the Cohen-Hoberman Inventory of Physical Symptoms (CHIPS) scale [2], which included energy impairment suggestive of a chronic health condition. Our choice to include disabling health conditions is consistent with classification systems and service provision contexts [24, 38]. This also helps address under-identification of disabling statuses; for example, about 25% of people with energy-impairing chronic conditions do not identify as disabled [15], but often still encounter similar barriers as people who do identify as disabled.

Analysis was guided by the research questions, largely focused on full sample portrayals of variable distributions as well as between-group tests of difference. We run both t-tests (parametric) and Mann-Whitney U tests (non-parametric) for significance testing; we had the same significance levels for all the pairwise comparisons, thus, we chose to report the results of t-tests for consistency. Most results meet $p < .01$ or better, and the number of repeat comparisons is small, so we did not correct for multiple comparison.

## 3.2 2020 Survey Participants

The 2020 study included 147 students (Table 2). There are 119 students without disabilities; of these, 48.7% (58) are female, 49.6% (59) are male and 1.7% (2) identified themselves outside of the male or female gender, 53.8% (64) are Asian, 11.8% (14) are underrepresented minorities (URMs: African or African-American, Alaska Native, Native American, Native Hawaiian Pacific Islander, or Latinx), 31.9% (38) are first-generation college students, and 15.1% (18) are LGBTQIA+. Only a few reported a pre-existing condition that made them vulnerable to COVID-19 (5 (4.2%)) and the average SES score is 5.9.

There are 28 students with disabilities/health concerns. Note that students could indicate multiple types of disability and that one student with a high CHIPS score also identified as disabled, which is why there are 24 disabled students and 5 students with high CHIPS scores. Students with disabilities/health concerns had higher marginalized status or vulnerabilities than students without in the study: 82.1% (23) are female, and 17.9% (5) male, 67.9% (19) are Asian, 39.3% (11) are first-generation college students, 21.4% (3) are LGBTQIA+ and the average SES is 5.5.

## 3.3 Repeat Participants

Although the majority of our analysis only looks at the 2020 survey participants, we also look at times at data collected in 2019. Of the 2020 students, 91 students returned from 2019 including 13 with disabilities/health concerns. These students filled out the same survey in 2019 and 2020, at the same time of year, with the exception of COVID-19 specific questions, which were added in 2020.

## 4 RESULTS

In this section, we first present students' levels of concern regarding the potential negative effects of the COVID-19 context on their educational standing and success. We then summarize and contrast students' adversity exposures, both those directly COVID-related as well as from sources such as discrimination, loneliness, and stress profiles via histories of major life adverse events. Finally, we compare students with disabilities/health concerns to those without disabilities/health concerns regarding their current mental health statuses.

| Demographics | Non-disabled | Disabled/in poor health |
|---|---|---|
| **Total sample** | **119** | **28** |
| % Female | 58 (48.7%) | 23 (82.1%) |
| % Asian | 64 (53.8%) | 19 (67.9%) |
| % URM | 14 (11.8%) | 1 (3.6%) |
| % FirstGen | 38 (31.9%) | 11 (39.3%) |
| % LGBTQIA+ | 18 (15.1%) | 3 (21.4%) |
| % Covid health vulnerability | 5 (4.2%) | 4 (14.3%) |
| Average SES | 5.9 | 5.5 |
| Vision or hearing impairment | | 14 (50%) |
| Mobility impairment | | 0 (0%) |
| Learning disability | | 2 (7.1%) |
| Mental health condition | | 8 (28.5%) |
| Other disability or impairment | | 1 (3.6%) |
| **Disabled** | | **24 (85.7%)** |
| **CHIPS [2]** | | 5 (17.9%) |

Table 2. Demographics. URM stands for under-represented minority (African-American, Latinx, Native American, and Pacific Islander); FirstGen stands for First Generation students (whose parents did not complete college). Covid health vulnerability are students who self-identified as having a pre-existing condition that puts them at risk for COIVD-19. Students self-identified their disability.

### 4.1 Educational Concerns

As noted in Table 1, the University of Washington courses went online approximately two weeks before the end of Winter Quarter (early March), and students filled out the survey after taking their last final. Part of that assessment was students' concerns about the potential impact of COVID-19 on their academic well-being. Figure 1 shows the percentages of students with and without disabilities/health concerns across eight indicators. At the time, things were in flux, but students knew that Spring quarter would be at least partly online. As is evident, concerns ranged high on many of these factors. More than 50% of students with disabilities/health concerns were very concerned about grades in both Winter and Spring quarters, as well as whether they would have to move degree requirements, negative impact on courses that could not then go online, admission to preferred major being threatened, and, to a lesser degree, negative impact on graduation and financial aid. Students with disabilities/health concerns were substantially more concerned ($M = 13.44, SD = 6.17$) than their non-disabled peers ($M = 8.25, SD = 6.28$), with means calculated as a sum across all items. This difference is significant ($t(145) = 3.94, p < .001$).

Given the wide range of disability identities represented in our sample, it is possible that not all students with disabilities/health concerns share the same levels of concern. For example, students with high fatigue may well prefer to stay home, whereas a student with hearing impairment may need to rapidly change how they access audio material through a transcription or sign language translation service. We anticipated that students with vision or hearing impairments might face greater barriers going online than those with mental health conditions or other disabilities/health concerns. However, of the 27 students who answered this question, we found that there was no significant difference ($t(26) = −0.23, p = .82$) for online classes concerns between the 13 students with vision or hearing impairments ($M = 21.2, SD = 5.37$) and the 14 students with other disabilities/health concerns ($M = 21.7, SD = 6.94$).

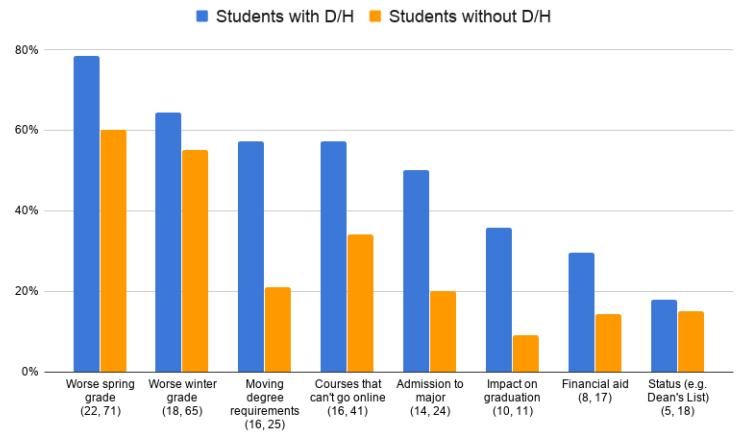

Fig. 1. Comparison of students with disabilities/health concerns (D/H) and those without who reported various educational concerns in the COVID-19 context. The X axis shows the type of concern. The Y axis shows percentage of students, within group. For example, 16 students with disabilities/health concerns were concerned about moving degree requirements (almost 60% of all students with disabilities) and 25 students without disabilities/health concerns (around 20%).

### 4.2 Adversity and Mental Health Trends

Toward assessing student vulnerabilities within the early COVID-19 context, we turn next to identifying areas of adversity exposure relevant to psychosocial functioning as well as mental health statuses as reflected through clinical measures.

*4.2.1 Adversity Sources.* We begin by investigating COVID-19 related stressors that students were experiencing outside of their educational settings and differences for those with disabilities/health concerns.

As can be seen in Figure 2, students with disabilities/health concerns experienced more stress than those without due to a range of financial, family and isolation issues. Experiencing tensions within the household during this period of isolation reflected the greatest disparity between students with disabilities/health concerns and their non-disabled peers. We also asked about whether social media use was a source of stress. This was high for all participants (over half agreed or strongly agreed); however the difference between students with and without disabilities/health concerns was not significant.

Table 3 reports group comparisons on four forms of adversity that could have implications for stress and functioning. In this data, the CEDH instrument measures chronic discrimination and harassment over the past twelve months, Major Life Events is capturing stressors over the lifetime, and COVID Related Adversities provides a sum of the stressors attributed specifically to COVID-19 (*i.e.* recent stressors) shown in Figure 2. Here we see that students with disabilities/health concerns report higher levels of chronic discrimination (*e.g.,* experiencing demeaning remarks or forms of unfair treatment) and higher exposure to major life adversities (*e.g.,* a serious interpersonal conflict, early life or recent maltreatment). Students with disabilities/health concerns also report more COVID-related concerns. Notably, both student groups reported comparable level concerns about isolation, and comparable loneliness.

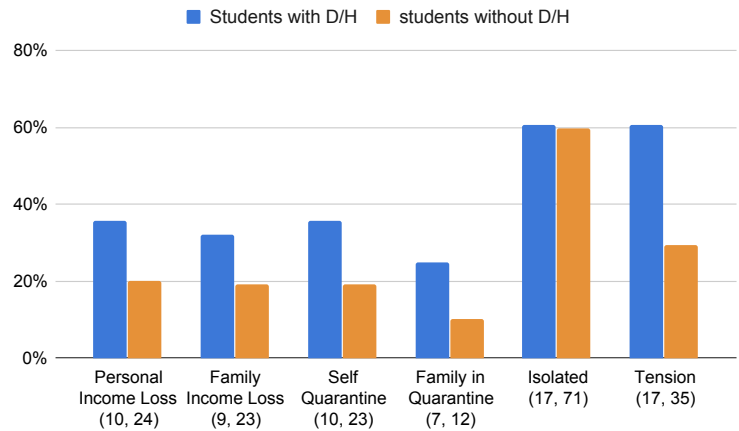

Fig. 2. Comparison of students experiencing stressful circumstances directly related to COVID-19 in early Spring 2020. The 6 circumstances assessed were major loss in own income, major loss in family income, oneself in quarantine, family member placed in quarantine, quarantine or other COVID-19 effect leading you to feel isolated, conflict/tension within household members due to COVID-19. The X axis lists relevant concerns, along with the number of students in each category. For example, in the case of *Income Loss*, 10 students with disabilities/health concerns (or 36% of those students) were concerned about income loss, while 24 students without disabilities/health concerns (20% of those students) were worried.

*4.2.2 Current Mental Health Status.* Given the greater levels of adversity exposures and variety that students with disabilities/health concerns face, we anticipated that students with disabilities/health concerns would experience increased mental health problems. However, both groups of students experienced comparable levels of current perceived stress, depression, and anxiety at the time they filled out the survey, as shown in Table 4.

It is notable that, in comparison to the other mental health measures, students with disabilities/health concerns did have significantly higher scores on the PTSD (distress) scale [43]. We delve more deeply into this result in the next subsection.

*4.2.3 Distress in students with disabilities.* To further explore these results, we undertook a more nuanced examination of student responses. Figure 3 shows the items which had the highest level of endorsement from students with disabilities/health concerns. These items are consistent with domains of PTSD reflective of general psychological distress (specifically, domains D and E [11]). That is, the items that were most frequently rated as highly concerning by students with disabilities/health concerns appear to be reflecting symptoms of hyperarousal and agitation–such as difficulty concentrating, trouble sleeping, and having negative feelings. Students were also comparatively cut off from others in this time of isolation, and indicated that as troubling.

| **Stress Exposure** | **D/H** | **No D/H** | **T test** |
|---|---|---|---|
| UCLA (Loneliness) [34] | 22.11 (5.87) | 22.24 (4.93) | -0.11 |
| CEDH (Chronic Discrimination) [26] | 11.75 (8.46) | 6.35 (6.47) | 3.17** |
| Major Life Event (general) | 13.25 (7.15) | 9.07 (4.22) | 2.98** |
| COVID-19 Related Adversities | 2.32 (1.79) | 1.42 (1.25) | 2.53* |

Table 3. Means (standard deviations) of scores on stress-related exposures for students with disabilities/health concerns (D/H) and without disabilities/health concerns (No D/H). T-test values and significance levels are indicated. Mann-Whitney U tests yielded comparable results. Significance is marked * $p < 0.05$, ** $p < 0.01$, *** $p < 0.001$.

| Mental Health Status | D/H | No D/H | T test |
|---|---|---|---|
| PSS (stress) [7] | 19.23 (6.9) | 18.88 (5.6) | 0.75 |
| CESD (depression) [32] | 9.00 (5.48) | 8.25 (4.75) | 0.67 |
| STAI (anxiety) [39] | 45.25 (10.48) | 44.14 (9.99) | 0.51 |
| PTSD (distress) [43] | 22.86 (15.99) | 13.18 (11.65) | 3.02** |

Table 4. Mean (standard deviation) of scores on mental health scales for students with disabilities/health concerns (D/H) and without disabilities/health concerns (No D/H). T-test values and significance levels are indicated. Mann-Whitney U tests yielded comparable results. Significance is marked as $*p < 0.05$, $**p < 0.01$, $***p < 0.001$.

*4.2.4 Returning Students.* One possible explanation for the PTSD group difference described above is that our sampling process in 2020 purposefully included more people with disabilities/health concerns and other vulnerabilities. To address this, we looked at the 2019 and 2020 data just for students who had participated in both the 2019 and 2020 studies (a total of 91 students, of whom 13 have disabilities/health concerns).

We first examined whether our 2020 results changed when we only looked at the 53 students who participated in both 2019 and 2020. Our findings confirm the significant increase in distress, with similar effect size (M=29.31 vs M=13.0 for D/H vs non D/H) and significance (p<.01). This difference existed in 2019 as well, but the effect size (M=17.5 vs M=8.18) and significance (p<.05) were both lower.

Not surprisingly (since it is a lifetime measure), the 13 returning students with disabilities/health concerns reported similar exposure to Major Life Events in both 2019 ($M = 11.15, SD = 5.18$) and 2020 ($M = 10.54, SD = 5.29$), $t(25) = 0.30, p = .77$. We also examined the change in PTSD scores from 2019 to 2020 for the 13 students with disabilities/health concerns. As Figure 4 shows, 10 of the 12 students who answered the PTSD scale reported higher levels of PTSD symptomology in 2020 than in 2019, the majority being 10 or more points higher in 2020, which bears clinical relevance. The difference in scores is also significant (2019 ($M = 17.5, SD = 13.57$), 2020 ($M = 29.31, SD = 16.13$), $t(23) = 44.0, p < .05$.

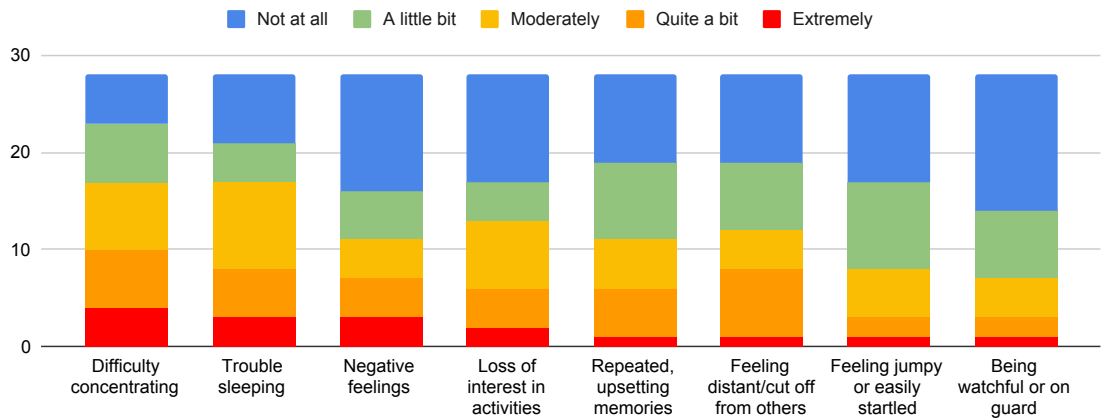

Fig. 3. Items in the PTSD scale endorsed by at least one person with disabilities/health concerns as extremely concerning

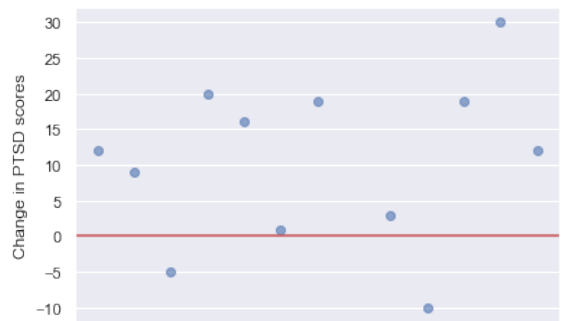

Fig. 4. Changes in PTSD scores of 12 returning students with disabilities/health concerns from last year to this year (12 of the 13 returning had responses for PTSD scale). Each dot represents a returning student, the y axis of each dot is the subtraction of PTSD scores from last year to this year. For example, if a student's PTSD scores were 16 and 5 in this year and last year, respectively, the value on the y-axis would be 11.

In comparison, for students without disabilities/health concerns, the change in score is smaller (only 19 out of 78 have an increase 10 points or more) and the difference in scores is significant in 2019 ($M = 8.18, SD = 9.6$) and 2020 ($M = 13.0, SD = 12.01$), $t(155) = -2.73, p < .01$. Note that while the significance is higher here, the means for students without disabilities/health concerns are lower than for students with disabilities/health concerns in both 2019 and 2020.

## 5 DISCUSSION

This study provides an early snapshot of the concerns and vulnerabilities of undergraduate students with disabilities/health concerns within the early-stage context of the COVID-19 pandemic. Such students expressed considerable concern about the educational and financial impacts of COVID-19. At this early stage of COVID-19, we did not see higher levels of self-reported depression, anxiety, stress or loneliness. However, students with disabilities/health concerns did report overall high levels of stress exposure and exposure to COVID-19 related adversities, and we saw evidence of associated distress.

### 5.1 Students with Disability/Health Concerns report more Stress Exposure and Higher Distress

By comparing students with and without disabilities/health concerns, we gain insights about the unique vulnerabilities of students with disabilities/health concerns in the COVID-19 context. They brought significantly deeper histories of both discrimination as well as serious major life adversity exposures than non-disabled students to this context. In addition, our results show that students with disabilities/health concerns are already experiencing COVID-19 related stress. Students reported higher or equal exposure to all of the COVID-19 specific stressors we measured than their peers without disabilities/health concerns. We also found early signs of distress, particularly difficulty concentrating, insomnia, and isolation, all understandable reactions to conditions of unpredictability and lack of control [44].

The timing of the surveys (early in the COVID-19 crisis) leaves open the possibility that the effects we see are due to different stress exposures between students with and without disabilities/health concerns and not specifically to COVID-19. However, our comparison of the students who provided survey data in both 2019 and 2020 suggests that exposure to discrimination and distress both increased this year and that our data reflect increased vulnerability of students with disabilities/health concerns in the COVID-19 context.

Prior evidence has established that students entering higher education with adversity backgrounds are at elevated risk of worsening mental health [19]. Subsequent stress exposure adds to and, in some cases, may exacerbate the effects of earlier adversities [14]. On the other hand, measures of stress, anxiety, depression, and loneliness all show no change (See Tables 3 and 4). This could also indicate resilience in the face of these stresses, many of which are areas where people with disabilities may have prior experience [6].

### 5.2 Students with Disability/Health Concerns need Better Online Education supports

Students with disabilities/health concerns in our study report worries about courses going online, and serious consequences of the move online including receiving worse grades, not being able to meet academic requirements online, and having admission to their chosen major impaired. In addition, our study shows that students are not only concerned about online courses, but also exposed to a variety of stressors from discrimination to financial concerns.

This suggests that more accessible online education tools should be designed to consider a wider range of concerns, which can potentially exacerbate the impact of barriers to access. Prior work has argued for the importance of considering context in the design of accessible technologies [33]. Rather than simply making online course materials more accessible, online instructors may need to support students who are coping with multiple stressors outside of class. For example, this might include designing for asynchronous participation, or redundant assessment that is robust to occasional absences. Further, in the context of a pandemic, best practices are to convey safety, calm, comfort and connectedness [40]. If educators and educational technology can strive to provide a calming, connecting experience, they can support students rather than compounding their stress.

### 5.3 Limitations

Our choice to combine people who identify as disabled with people who report themselves as having significant health concerns is driven by the relatively high numbers of people with health concerns who do not identify as disabled [15], even though they may experience accessibility barriers. Moreover, this group comprised fewer than 20% of the students with disabilities/health concerns and in an analysis comparing subgroups of students with disabilities/health concerns, we did not see major differences in our results. There are philosophical questions raised by this choice about who "counts" as disabled. Our view was that, relative to our study's goals, this level of health concern rendered these students comparably vulnerable in the COVID-19 context. That said, we recognize that other sources may use differing definitions.

Finally, we note that we have a relatively small and non-representative sample (for either Seattle or nationally with respect to undergraduates). This sample reflects our efforts to reach more vulnerable students and thus our results may be most relevant to students with such vulnerabilities.

## 6 CONCLUSION AND FUTURE WORK

This article is a first look at the impact of COVID-19 on students with disabilities, and as such, it illuminates a range of needs not yet well documented. It also helps generate questions for further inquiry.

Understanding how COVID-19 affects students with disabilities/health concerns can help us to identify patterns of concerns, stress and indicators of mental health, and provide further steps to take or consider in order to respond to an unexpected global health pandemic. Our findings show that students with disabilities/health concerns were more worried about the outcomes of the unanticipated change to online learning than their peers. In addition, during our study, students with disabilities/health concerns experienced more COVID-19 related adversities and distress.

As pandemics (as well as other major stress conditions) continue over time, additional stressors begin to arise (e.g., loss of income, lack of healthcare access, caring for others [30, 37]). This could in turn translate to larger impacts on the wellbeing of students with disabilities/health concerns. In the future, we hope to compare students' response to COVID-19 further into the epidemic to the data described in this paper. Further, we plan to conduct interviews to develop a more nuanced, qualitative picture of these impacts.

To conclude, we argue that just as students with disabilities belong in higher education, accessibility research belongs in higher education. It is time for the accessibility community to expand our work studying technology use and developing technology to support students with disabilities in higher education. We hope that these insights into the student experience demonstrate how important it is to learn how higher education fails and succeeds to support students with disabilities, and where technology can play a role in improving this.

## ACKNOWLEDGEMENT

This material is based upon work supported by the National Science Foundation under Grant No. EDA-2009977, CHS-2016365, and CHS-1941537. Funding was also provided through a grant from Samsung and a Google Security and Privacy unrestricted gift.